\documentclass[aps,superscriptaddress,twocolumn,a4paper]{revtex4}
\usepackage{graphicx}
\usepackage[all]{xy}
\usepackage{amsmath}
\usepackage{amssymb}
\usepackage{tensor}
\usepackage{color}
\usepackage[dvips]{epsfig}
\usepackage[dvips]{graphicx}
\newcommand{\be}{\begin{equation}}
\newcommand{\ee}{\end{equation}}
\newcommand{\ben}{\begin{eqnarray}}
\newcommand{\een}{\end{eqnarray}}
\newcommand{\bes}{\begin{subequations}}
\newcommand{\ees}{\end{subequations}}
\newcommand{\bb}{\bibitem}
\def\bal#1\eal{\begin{align}#1\end{align}}

\newcommand{\bfi}{\begin{figure}}
\newcommand{\efi}{\end{figure}}
\newcommand{\bc}{\begin{center}}
\newcommand{\ec}{\end{center}}

\begin{document}
\title{Geometrically constrained localized configurations engendering non-topological profile}

\author{D. Bazeia{\footnote{Corresponding author, email address: bazeia@fisica.ufpb.br\\https://orcid.org/0000-0003-1335-3705}}}
\address{Departamento de Física, Universidade Federal da Paraíba, João Pessoa, Paraíba, Brazil}

\author{Isaiane Bezerra}
\address{Departamento de Física, Universidade Federal da Paraíba, João Pessoa, Paraíba, Brazil}

\author{R. Menezes}
\address{Departamento de Ciências Exatas, Universidade Federal da Paraíba, Rio Tinto, Paraíba, Brazil}
\address{Unidade Acad\^emica de F\'{\i}sica, Universidade Federal de Campina Grande, Campina Grande, Para\'{\i}ba, Brazil}

\begin{abstract}
This work deals with two real scalar fields in two-dimensional spacetime, with the fields coupled to allow the study of localized configurations. We consider models constructed to engender geometric constrictions, and use them to investigate solutions of the lump type, which attain no topological properties. We show how to modify the internal structure of the field configurations and the corresponding energy densities in several distinct ways, making them thinner, thicker and also, in the form of a multi-lump solution composed of two or more lumps or asymmetrically distributed around their associated centers. The results appear to be of current interest in high energy physics and may, in particular, be used to study bright solitons in optical fibers and in Bose-Einstein condensates.
\end{abstract}

\maketitle

{\bf Introduction. --} In high energy physics, topological structures such as kinks, vortices and magnetic monopoles are important localized configurations that attain topological properties, which in general protect them against instability \cite{B1,B2}. However, there are other localized structures; some of them, in particular, do not engender topological properties, so they need different mechanisms to provide stability. Typical possibilities are q-balls, which are charged localized configurations that attain stability due to other characteristics, such as the charge associated to them; see, e.g., Refs. \cite{Q0,Q1,Q2,Q3,Q4,Q5} and references therein for more information on q-balls. 
The simplest q-balls appear from complex scalar field in two-dimensional spacetime, but they may engender other features, as gauged q-balls in the presence of gauge fields and as boson stars when minimally coupled to gravity, as investigated in \cite{Ga1,Ga2,B} and in references therein.

The present work is concerned with lumplike configurations, which are described by real scalar fields engendering localized solutions with finite energy and no topological properties. They have the profile of a bell-shaped configuration, and although they are in general unstable in high energy physics, they have the same bell-shaped form of the soliton that appear in the Korteveg-de Vries equation, one of the first nonlinear differential equations used to describe solitary wave in shallow water, where nonlinearity and dispersion conspire to give rise to soliton; see, e.g., Refs. \cite{ref1,ref2,ref3,ref4,ref5} and references therein for further information on solitons. Lumplike solutions can also be mapped into bright solitons, which play an important role as optical solitons in fibers \cite{OF} and into matter waves in Bose-Einstein condensates \cite{MW,BE}. In particular, one can consider the nonlinear Schrödinger equation to construct explicit nontrivial resonant bell-shaped solutions, with the trapping potential and nonlinearities depending both on time and on the spatial coordinates \cite{Kono,abc}.

Scalar fields are also of importance in high energy physics in other contexts, for instance, to describe the dilaton, which may contribute to the holographic descriptions of the strongly coupled quark-gluon plasma produced in relativistic heavy-ion collisions; see, e.g., Ref. \cite{RR} and references therein. They may also be considered as axions, which can play interesting role related to studies in string theory, considered as plausible solution of the strong CP problem \cite{ST} and in cosmology and astrophysics \cite{CO}, in particular, as a dark matter candidate. Lumplike configurations may also be known as sphalerons, which have gained increasing interest recently; see, e.g., Refs. \cite{spha1,spha2,spha3} and references therein.   

Due to the several possibilities to study scalar fields and to use bell-shaped lumplike configurations in distinct areas of nonlinear science, we think they deserve being further studied in high energy physics as well, where they may appear in relativistic models described by real scalar fields in 1+1 spacetime dimensions. In order to bring novelties, here we consider models described by two real scalar fields, $\phi$ and $\chi$, owing to use the second field, $\chi$, to geometrically constrains the first field, $\phi$, to change its internal structure. This possibility was first considered in \cite{G1}, and it was further investigated in Refs. \cite{G2,G3,G4,G5} in several distinct scenarios, considering that both the $\phi$ and $\chi$ fields support kinklike configurations. In the present work, however, we are interested to study another possibility, in which the field $\phi$ gives rise to lumplike configurations, aiming to use the second field $\chi$ to induce modifications in the internal structure of the lumplike solutions.

As one knows, when compared with kinks, the study of lumps is more complicated, because they in general do not obey first-order differential equations, so it is much harder to find analytical solutions for lumps. However, in this work we shall follow the lines of Ref. \cite{p2}, which provides an interesting way to describe lumplike configurations that obey first-order differential equations and, in this sense, makes it easier to investigate lumps in high energy physics. 
With this motivation, we have organized this work as follows: In the next section we deal with the methodology, that is, we describe the main steps one needs to present the model, write the equations of motion, describe the energy density and introduce the first-order equations that solve the equations of motion and minimize the total energy of the solutions. We then go on and illustrate the main results investigating several distinct specific possibilities, showing how each model works to modify the internal structure of the corresponding lump. The work is ended with a breaf summary of the results, and the inclusion of comments on distinct possibilities of continuation of the present study.

{\bf Methodology. --} We start working with two real scalar fields, $\phi$ and $\chi$, and consider the two-field model described by the Lagrange density
 \begin{equation} \label{eq: lagrangiana modificada}
        \mathcal{L} = \frac{1}{2}f(\chi)\partial_\mu\phi\partial^\mu\phi + \frac{1}{2}\partial_\mu\chi\partial^\mu\chi - U(\phi,\chi),
    \end{equation}
where $f(\chi)$ is a non-negative function of the field $\chi$ and $U(\phi,\chi)$ is the potential. Here we are using natural units, and considering dimensionless variables to describe the fields and the space and time coordinates. Also, the metric is such that $x^\mu=(t,x)$ and $x_\mu=(t,-x)$. The equations of motion are given by
\begin{eqnarray}
        \partial_\mu \left( \frac{\partial \mathcal{L}}{\partial (\partial_\mu \phi)} \right) + \frac{\partial{U}}{\partial \phi} = 0, \label{eq: eq movimento modificadas} \\
        \partial_\mu \left( \frac{\partial \mathcal{L}}{\partial (\partial_\mu \chi)}  \right) -\frac12 \partial_\mu\phi\partial^\mu\phi \frac{df(\chi)}{d\chi} +\frac{\partial {U}}{\partial \chi} = 0,
    \end{eqnarray}   
and for static solutions they become
\begin{eqnarray}
        \frac{d}{dx}\left( f(\chi)\frac{d\phi}{dx}\right) -\frac{\partial U}{\partial\phi}=0, \label{eq: movimento modificada 1} \\
          \frac{d^2\chi}{dx^2} - \frac{1}{2}\frac{df(\chi)}{d\chi}\left( \frac{d\phi}{dx}\right) ^2 - \frac{\partial U}{\partial \chi}=0. \label{eq: movimento modificada 2}
    \end{eqnarray} 
Moreover, the energy density is
\begin{equation} \label{enedes}
        \rho = \frac{1}{2} f(\chi) \left( \frac{d\phi}{dx}\right)^2 + \frac{1}{2} \left( \frac{d\chi}{dx}\right)^2 + U(\phi, \chi).
    \end{equation}

In order to study how a kinklike structure can geometrically constrain the profile of a lumplike structure, we then suppose that the $\chi$ field has a kink profile. Here we consider the case of a standard kink, described by the solution
\be
\chi(x)=\tanh(\alpha\, x),
\ee
where $\alpha$ is a real parameter that controls the width of the kink, which is centred at the origin of the spatial coordinate. To achieve this possibility, we take the potential in the form
\be   
U(\phi,\chi)=\frac{V(\phi)}{f(\chi)}+ \Gamma(\chi),
\ee
where $\Gamma(\chi)$ is written in terms of another function, $\Omega(\chi)$, in the form
\be
\Gamma(\chi)=\frac12\left(\frac{d\Omega}{d\chi}\right)^2,
\ee
and we take
\be
\Omega(\chi)=\alpha\chi-\frac{\alpha}{3}\chi^3.
\ee
This allows that we write the energy density \eqref{enedes} in the form $\rho=\rho_1+\rho_2$, where $\rho_2$ depends only on the field $\chi$, and has the form
\be
\rho_2= {\alpha}^2 {\rm{sech}}^4 (\alpha x).
\ee
Also, $\rho_1$ is given by
\be
\rho_1=\frac12 f(\chi)\left( \frac{d\phi}{dx}\right)^2 + \frac{V(\phi)}{f(\chi)}.
\ee

Now, we have to pay attention to the fact that lumps requires that the potential has to cross the zero, passing from positive to negative values, so its square root cannot be always real. To overcome this problem, we follow Ref. \cite{p2} and think of a lump as composed by a kink joined together with an antikink at the origin, so we can write a first-order framework for lumps, splitting the real line into two distinct portions, one for negative values of $x$, and the other for positive values. This means that we can introduce an auxiliary function $W=W(\phi)$ and write first-order equations for the field $\phi$ in the form
\be\label{fo1}
\frac{d\phi}{dx}=\pm\frac{W_\phi}{f(\chi)},\;\;\; {\rm for}\;\;\; x<0,
\ee
and 
\be\label{fo2}
\frac{d\phi}{dx}=\mp\frac{W_\phi}{f(\chi)},\;\;\; {\rm for}\;\;\; x>0,
\ee
where $W_\phi$ stands for the derivative of $W$ with respect to the field $\phi$, that is, $W_\phi=dW/d\phi$. In this case, the potential $V(\phi)$ is written as 
\be
V(\phi)=\frac12 {W^2_\phi},
\ee
and the energy density $\rho_1$ becomes
\be
\rho_1=\frac12f(\chi)\left(\frac{d\phi}{dx}\mp\frac{W_\phi}{f(\chi)}\right)^2\pm\frac{dW}{dx},
\ee
or
\be
\rho_1=\pm\frac{dW}{dx},
\ee
when the first-order equations \eqref{fo1} and \eqref{fo2} are satisfied. 

The above results were obtained on general grounds, so they can be used to describe specific models, to investigate how the kinklike configuration of the field $\chi(x)=\tanh(\alpha x)$ can contribute to modify the lumplike structure to be described by the field $\phi$. This issue will be explored in the section below.

\begin{figure}
\includegraphics[width=0.23\textwidth]{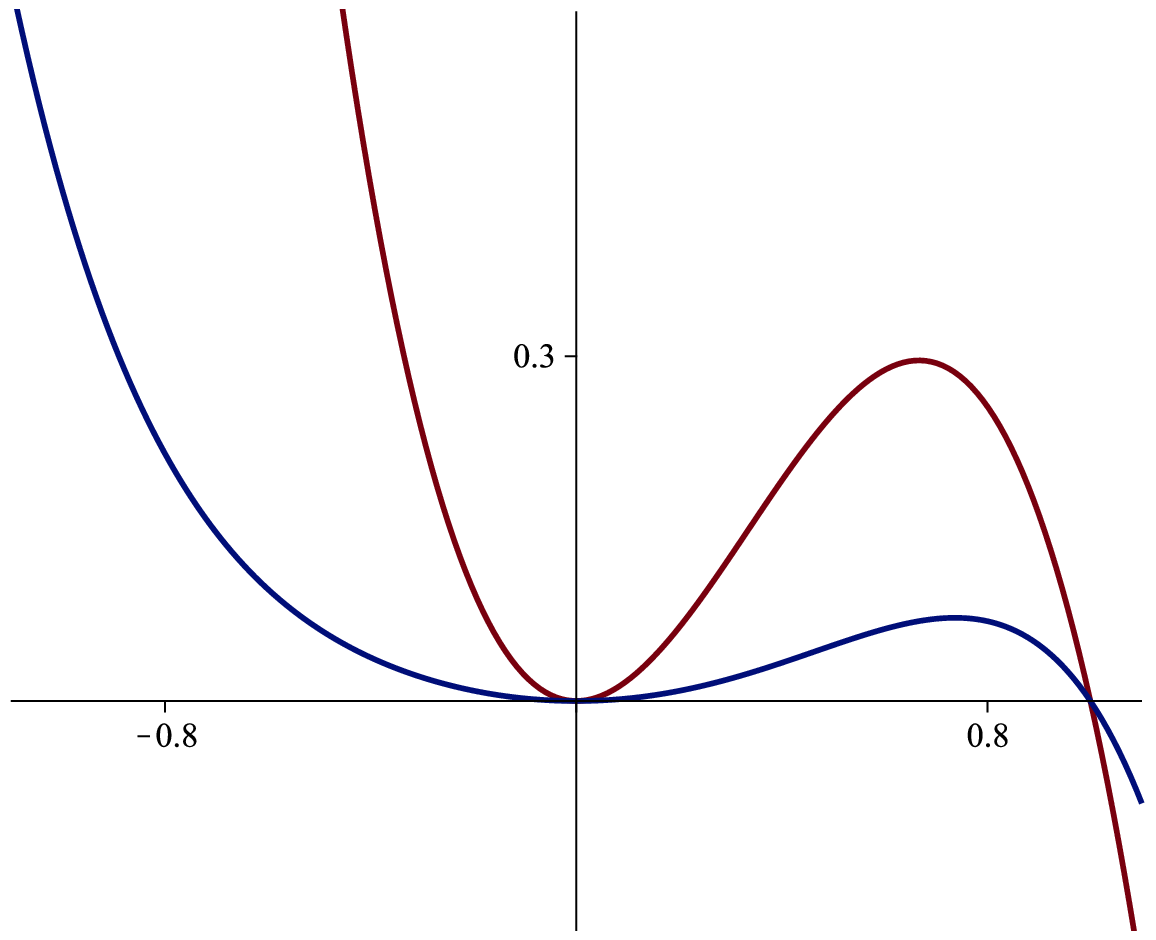}
\includegraphics[width=0.23\textwidth]{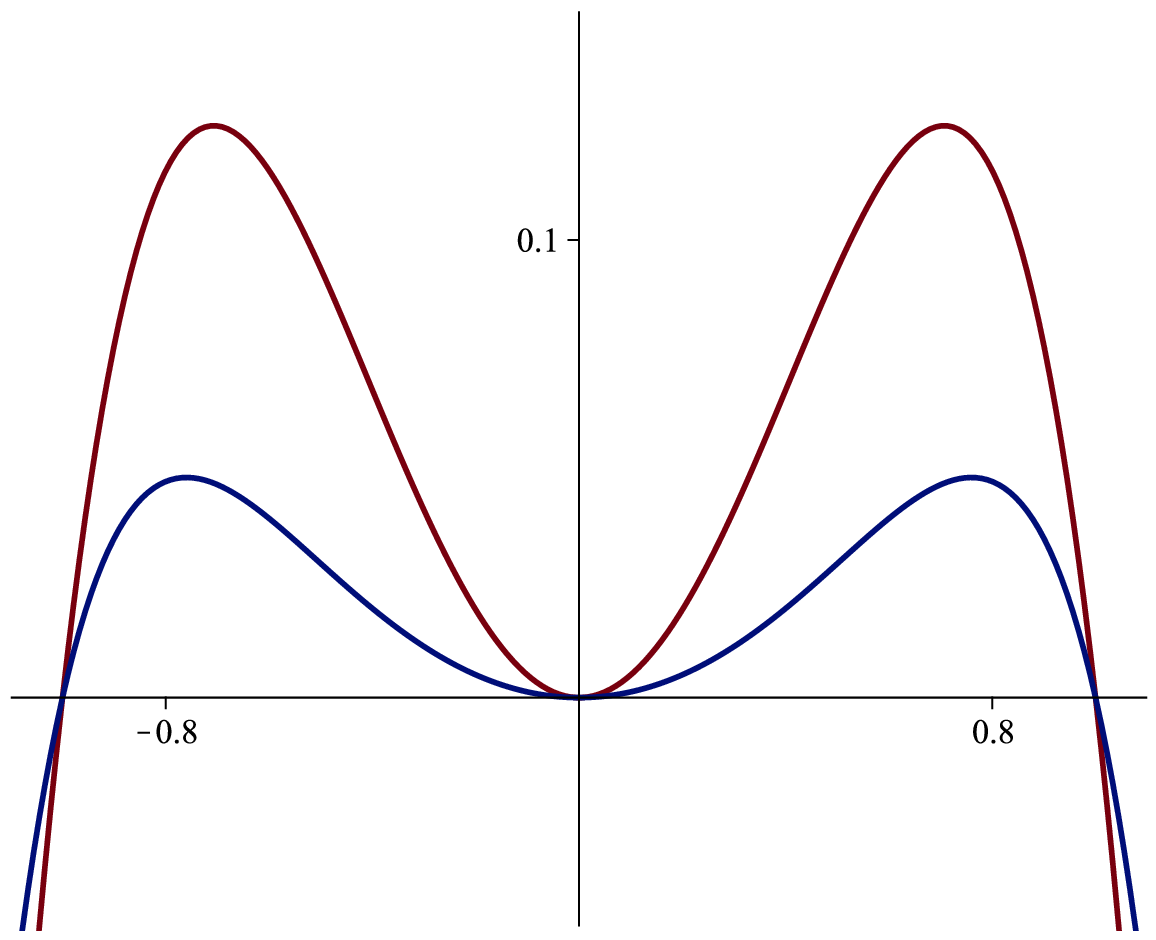}
\includegraphics[width=0.23\textwidth]{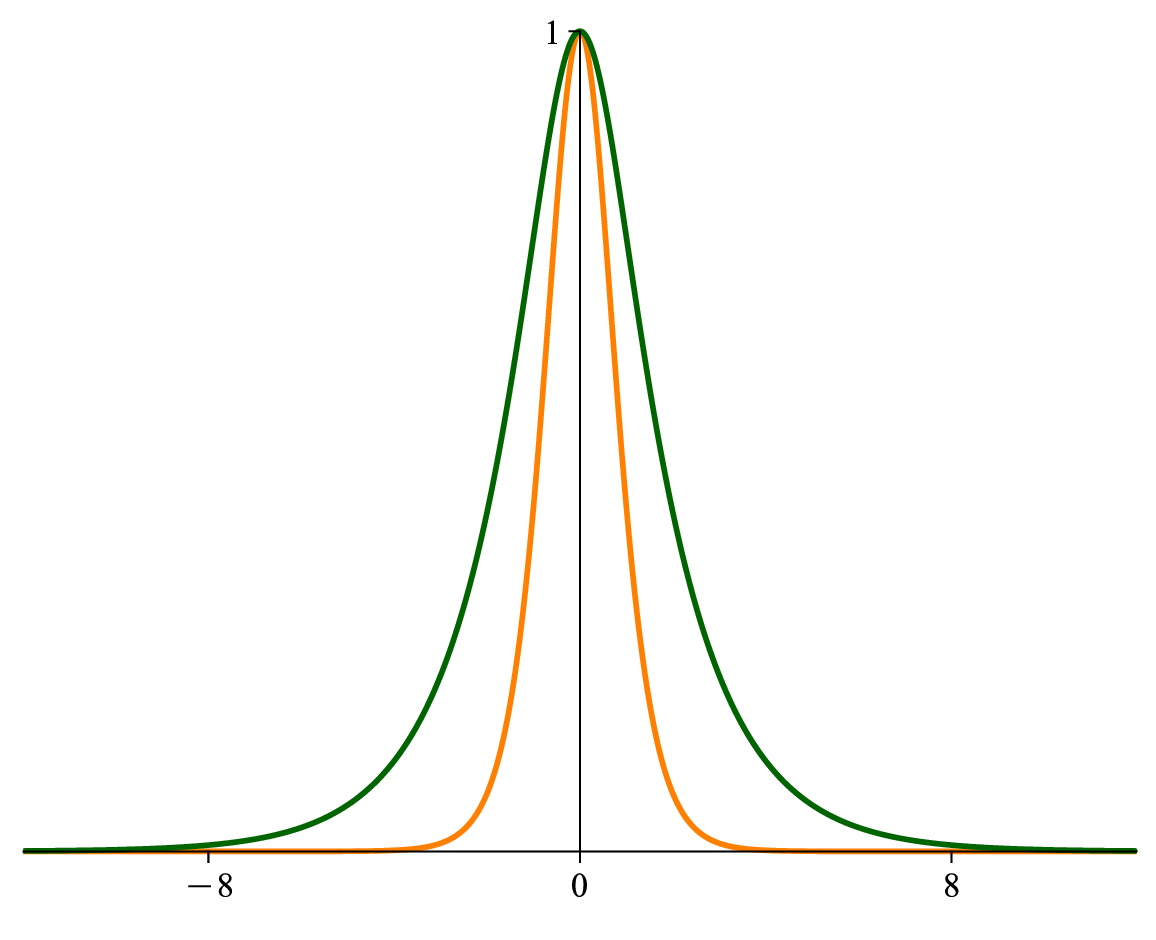}
\includegraphics[width=0.23\textwidth]{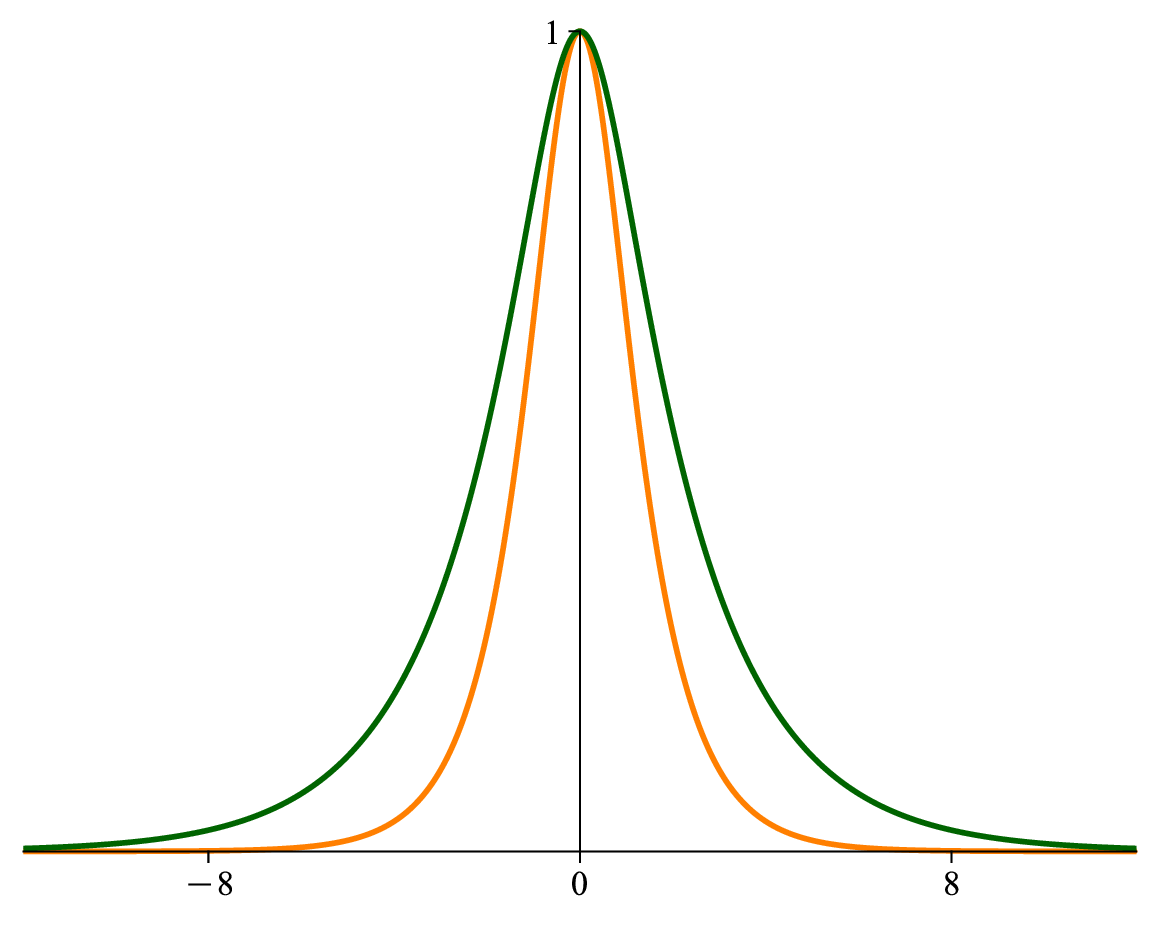}
\caption{Top panel: The potential $V(\phi)$ of Eq. \eqref{pot01} as a function of $\phi$, depicted for $n$ odd (left)  and even (right) with $n=1$ (red) and $3$ (blue), and with $n=2$ (red) and $4$ (blue). Bottom panel: The lump solution $\phi(x)$ of Eq. \eqref{lump01} as a function of $x$, also depicted for $n$ odd (left) and even (right), with $n=1$ (orange) and $3$ (green), and with $n=2$ (orange) and $4$ (green).}
\label{fig1}
\end{figure}

{\bf Illustrations. --} Let us now use the above general results to construct explicit examples of distinct lumplike solutions, as we describe below. We consider the model which supports lumplike structure being described by the family of polynomial potentials
\be\label{pot01}
V_n(\phi)=\frac{2}{n^2}\, \phi^2\,(1-\phi^n),
\ee
where $n=1,2,...$ is positive integer. These potentials were introduced in Ref. \cite{Wesley}, with the corresponding lumplike solutions 
\be\label{lump01}
\phi_n(x)= {\rm sech}^{2/n}(x).
\ee
We notice that these potentials can be separated into two distinct families, one for odd values on $n$, and the other for even values. For this reason, in Fig. \ref{fig1} we depict the potential and the lump solution for $n=1,3$ and for $n=2,4$ separately.

We can use these results to start exploring the idea developed in the previous section. Toward this goal, let us write 
\be
W_\phi=\frac{2}{n}\, \phi \,(1-\phi^n)^{1/2}.
\ee
We see that $W_\phi$ may not be real when $\phi$ is outside the interval $[-1,1]$. To avoid problem with this, we shall then search for lumplike solutions such that $ \phi(x)\in [-1,1]$, and below we explore several possibilities, controlled by some specific choices of the function $f(\chi)$.

\begin{figure}
\includegraphics[width=0.23\textwidth]{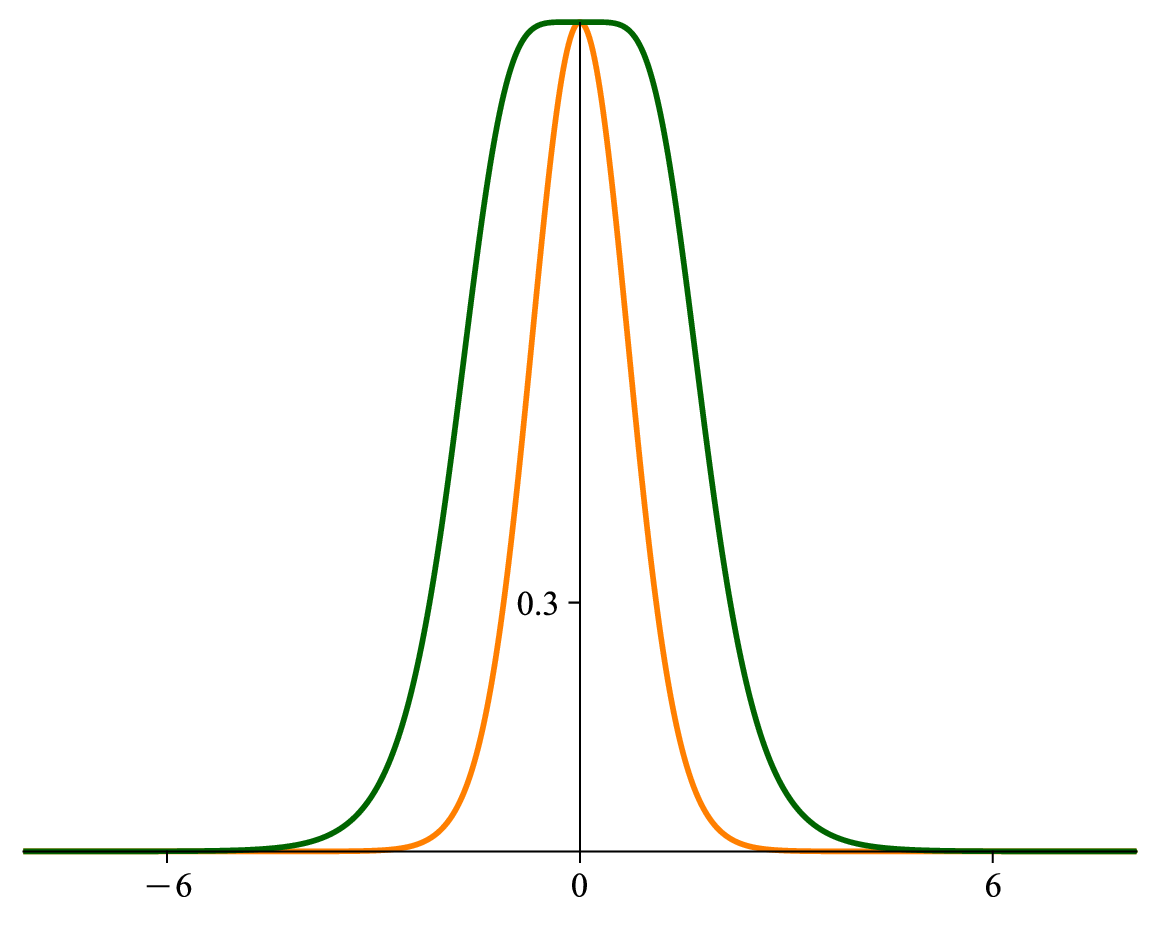}
\includegraphics[width=0.23\textwidth]{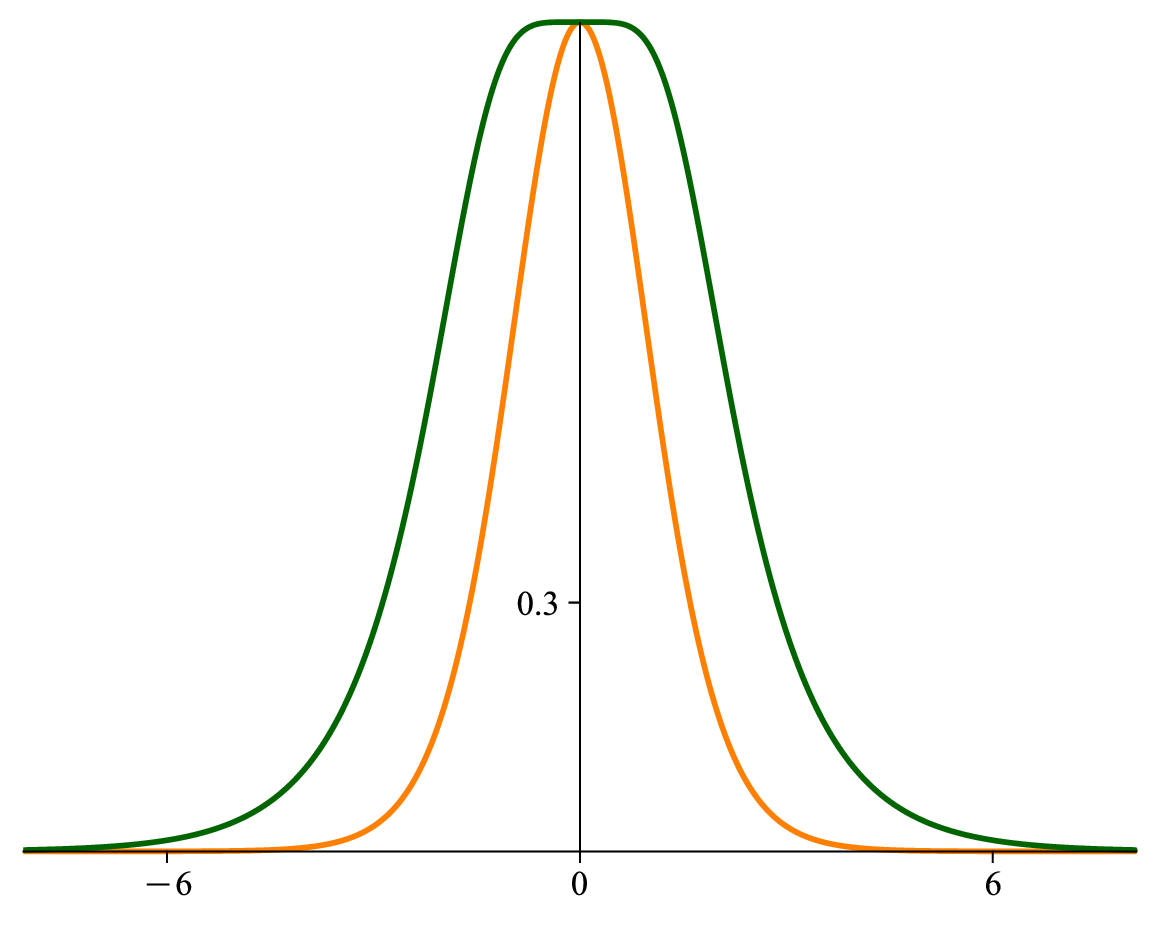}
\includegraphics[width=0.23\textwidth]{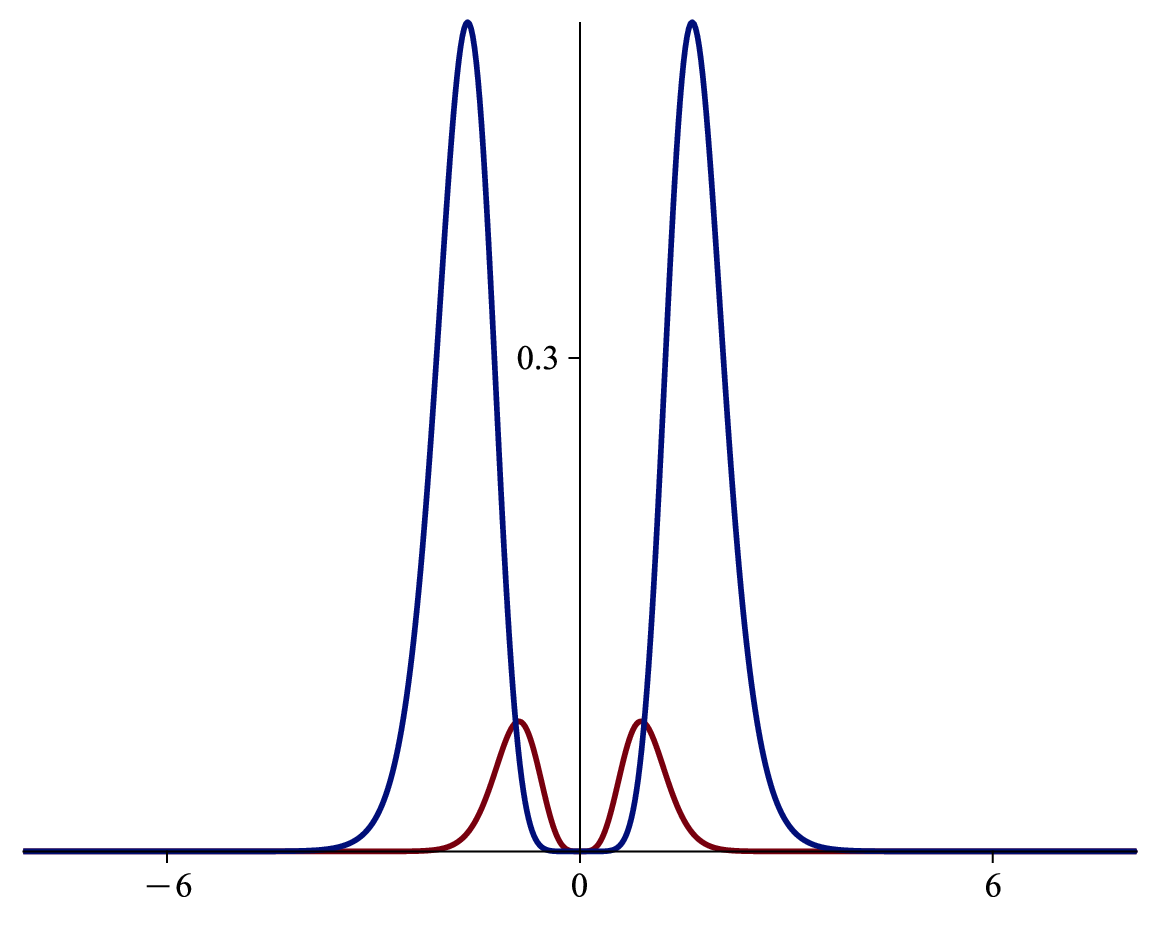}
\includegraphics[width=0.23\textwidth]{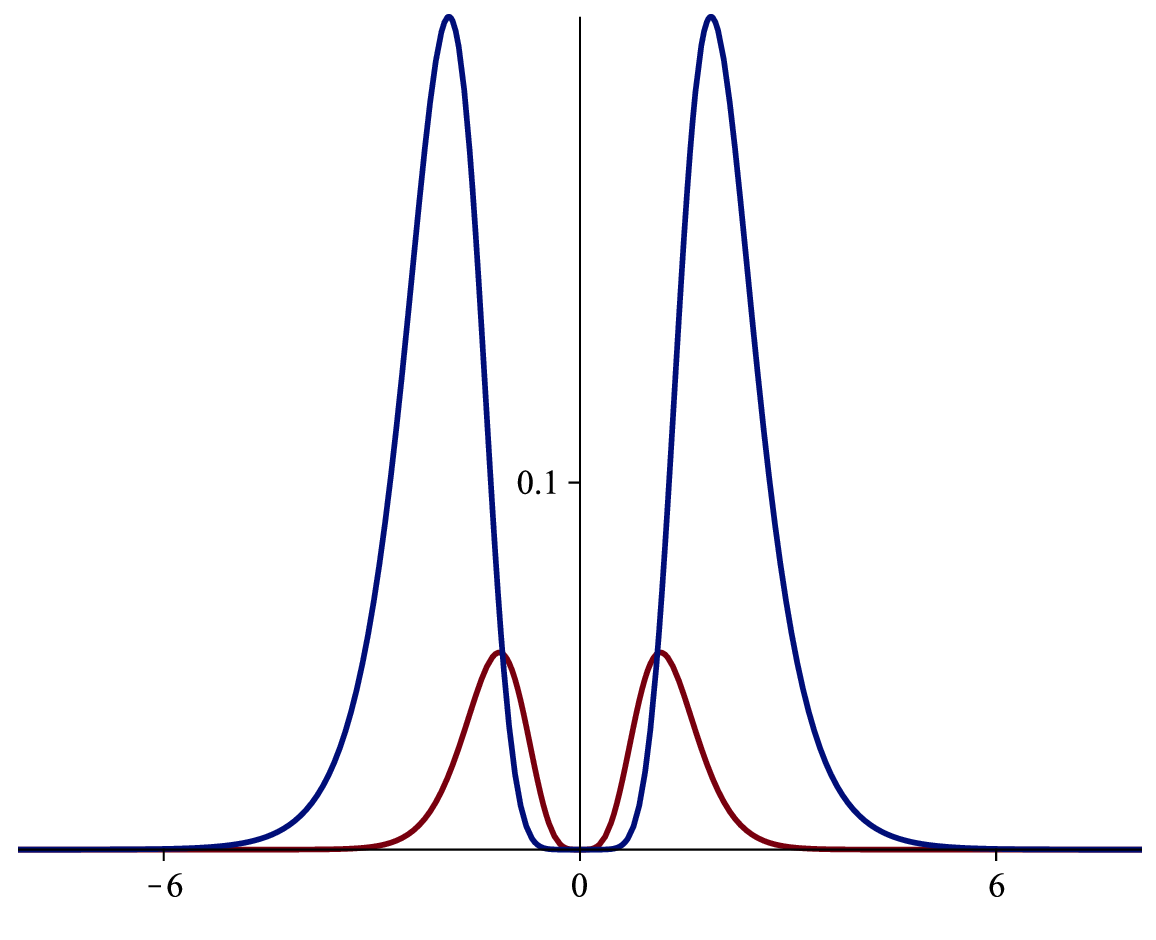}
\caption{Top panel: The lump solution $\phi(x)$ of Eq. \eqref{chi2} as a function of $x$, depicted for $n=1$ (left) and $2$ (right), with $\alpha=0.5$ (orange) and $1.0$ (green).
 Bottom panel: The energy density $\rho_1(x)$ in Eq. \eqref{ene1} as a function of $x$, depicted for $n=1$ (left) and $2$ (right), for $\alpha=0.5$ (red) and $1.0$ (blue).}
\label{fig2}
\end{figure}

{\it A. First model. --} To describe an interesting possibility, let us choose $f(\chi)=1/\chi^2$. This was first considered in \cite{G1} in the case of kinks, and shown to add an internal modification similar to the geometric constriction described before in \cite{M1} in the study of magnetic materials. In the model described in \cite{G1}, the kinklike solution acquires a profile similar to the two-kink structure described before in Refs. \cite{CL,BMM}, so our motivation here is to show how it works for lumps. In this case, a possible solution of the first-order equations \eqref{fo1} and \eqref{fo2} can be written as
\be\label{chi2}
\phi(x)={\rm sech}^{2/n}(\xi(x)),
\ee
where
\be
\xi(x)=\frac1{\alpha}\,(x-\tanh(\alpha\,x).
\ee
This is a new lumplike configuration which we depict in Fig. \ref{fig2} for $n=1,2$ and for $\alpha=0.5, 1.0$. We see that the geometric constriction introduced by the kink $\chi(x)$ and the function $f(\chi)=1/\chi^2$ changes importantly the profile of the bell-shaped lumplike solution, making it thinner or thicker, depending on the choice of $n$ and $\alpha$. Here it is also possible to identify the contribution $\rho_1$ for the energy density, which gets to the form
    \begin{equation} \label{ene1}
        \rho_1(x) = \frac4{n^2}\tanh^2(\alpha x) \tanh^2(\xi(x)) {\rm {sech}}^{4/n}(\xi(x)).
    \end{equation}
It is also depicted in Fig. 2 for some values of $n$ and $\alpha$, and also changes significantly as $n$ and $\alpha$ change.

{\it B. Second model. --} In order to further illustrate how the geometric constriction works to modify the standard lump, let us consider 
\be 
f(\chi)=\frac{1}{\cos^2(m \pi \chi)},
\ee
with $m$ being a natural number. The motivation here is to show how the periodic property of the cosine works to change the internal profile of the lumplike configuration. In this case, the field configuration has the form

\be\label{cos}
        \phi(x) = {\rm sech}^{2/n} (\eta (x)),
        \ee
        where
        \be
        \eta(x) = \frac{1}{2}\,x + \frac{1}{4 \alpha}\;({\rm Ci}(\xi^+_m(x)) - {\rm Ci}(\xi^-_m(x))), \ee
        and
        \be
         \xi^{\pm}_m (x) = 2\pi m(1 \pm \tanh(\alpha x)).
\ee 
Here the ${\rm Ci}(y)$ is the cosine integral, and the energy density $\rho_1$ is given by
 \be\label{ene2}
        \rho_1(x)\! =\!\frac4{n^2} \!\cos^2(m\pi \tanh(\alpha x))\, {\rm sech}^{4/n}(\eta(x))\tanh^2(\eta(x)).
\ee
We depict in Fig. 3 the solution \eqref{cos} and the above energy density $\rho_1$ for some values of the parameters involved in the model. Interestingly, the solution acquires internal structure which can be enriched by distinct choices of  the parameter $m$. In Fig. 3, we depict the case of $m=1$, which shows a solution having a structure composed of a lump on top of another lump, in the form of a two-lump configuration. For $m=2$ we would have another composition, in the form of a three-lump solution, and so on. In this sense, the parameter $m$ identifies a multi-lump configuration, which can be used in application of practical use. 

\begin{figure}
\includegraphics[width=0.23\textwidth]{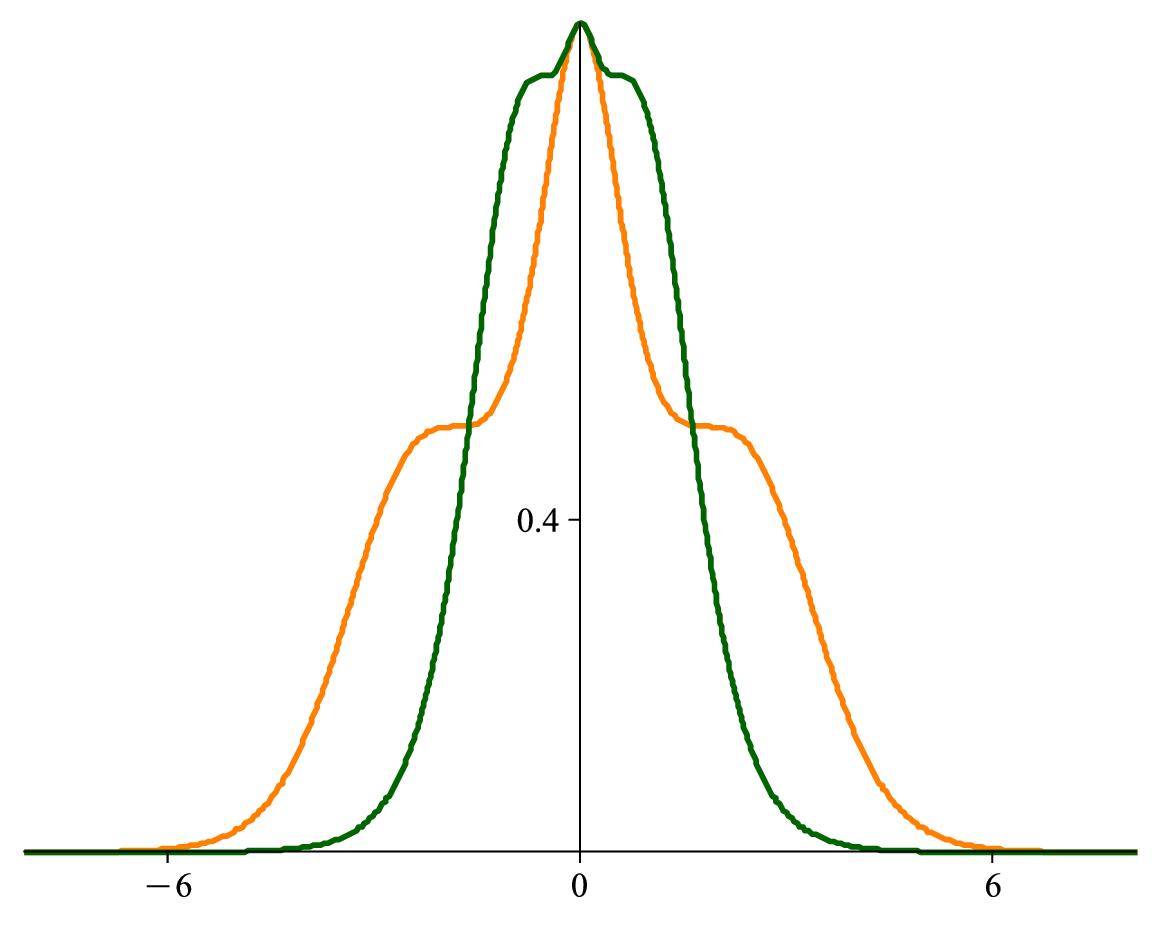}
\includegraphics[width=0.23\textwidth]{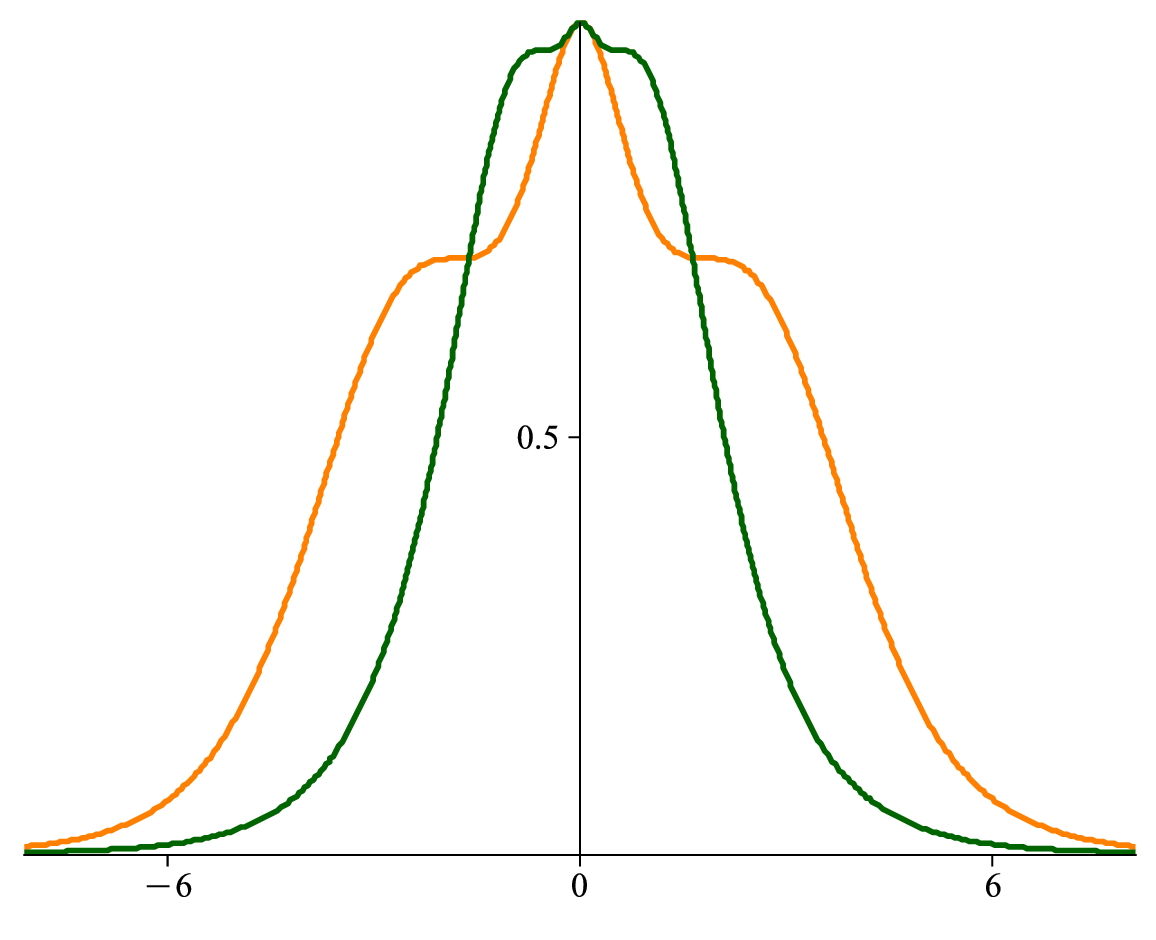}
\includegraphics[width=0.23\textwidth]{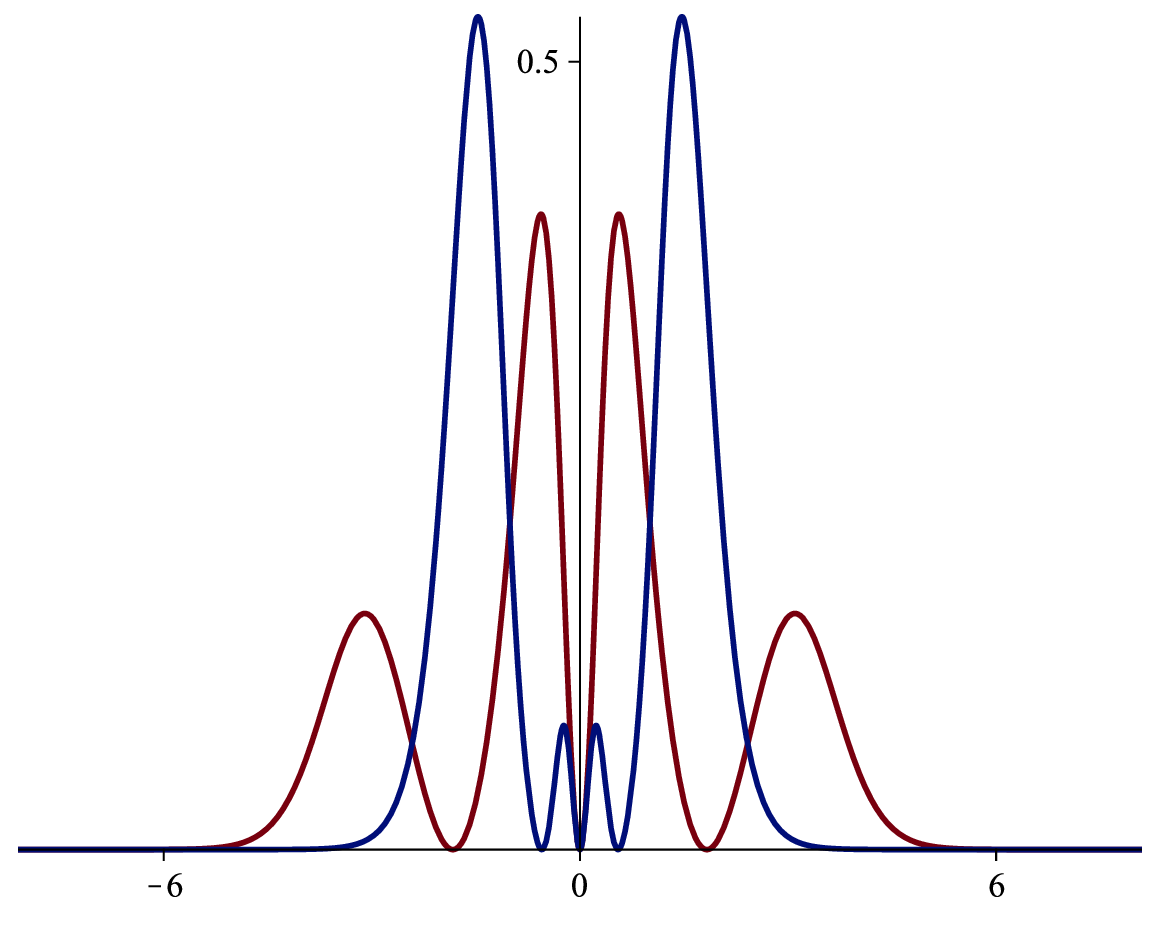}
\includegraphics[width=0.23\textwidth]{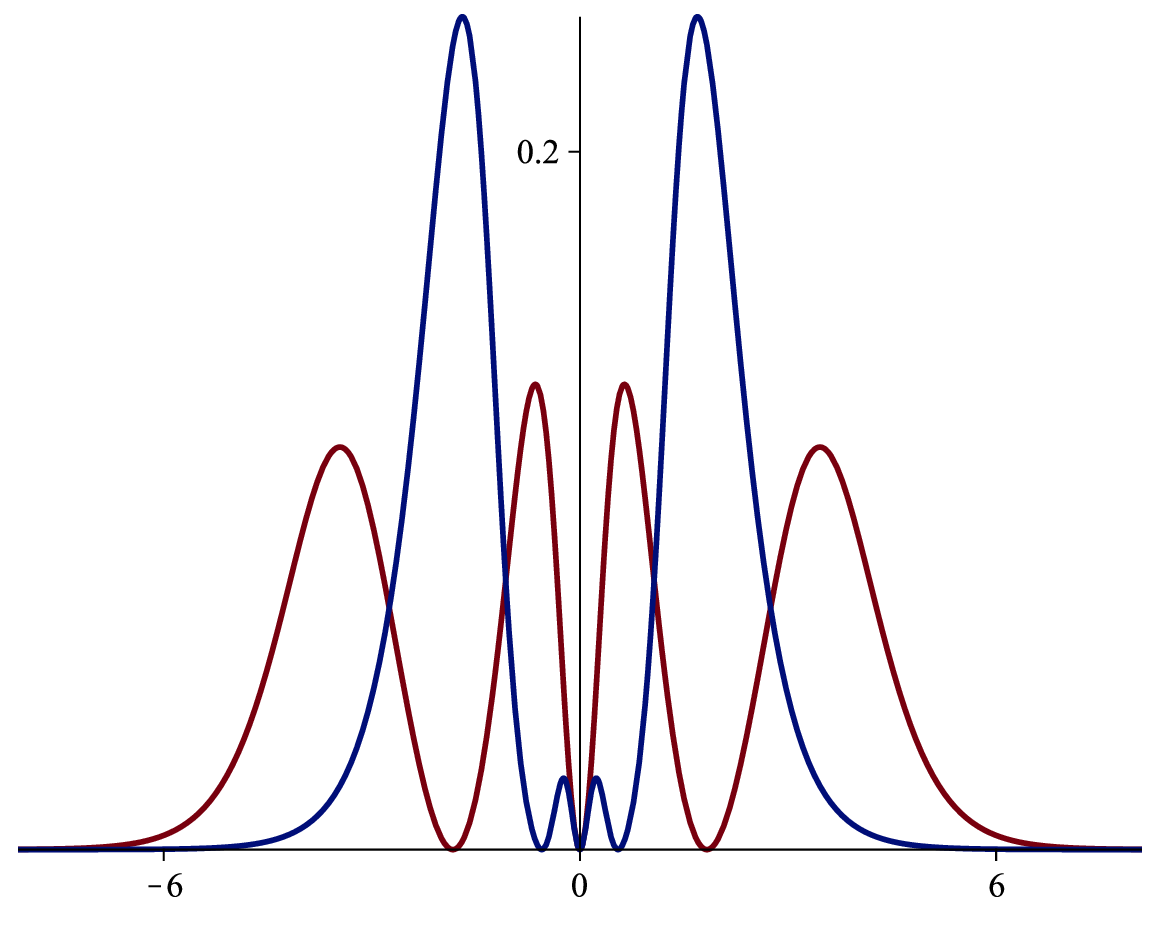}
\caption{Top panel: The lumplike solution $\phi(x)$ in Eq. \eqref{cos}, depicted for $n=1$ (left) and $2$ (right), for $\alpha=0.3$ (orange), and $1.0$ (green), and for $m=1$. Bottom panel: The energy density $\rho_1(x)$ in Eq. \eqref{ene2}, depicted for $n=1$ (left) and $2$ (right), for $\alpha=0.3$ (red) and $1.0$ ( blue), and for $m=1$.}
\end{figure}

{\it C. Third model. --} Another model is constructed using $f(\chi)$ in the form
\be
f(\chi) = \frac{1}{\sin^2((m + \frac{1}{2})\pi \chi)},
    \ee
with $m$ being a natural number. In this case the solution is

\be\label{sin}
        \phi(x) = {\rm sech} ^{2/n}(\eta(x)), 
\ee
where
\be\label{sol3} 
\eta(x) = \frac{1}{2}x +
\frac{1}{4 \alpha}({\rm Ci}(\xi^+_m(x))- {\rm Ci} (\xi^-_m(x))),
\ee
and
\be
 \xi^{\pm}_m (x) = (2m+1)\pi (1\pm \tanh(\alpha x)).
\ee
        
\begin{figure}
\includegraphics[width=0.23\textwidth]{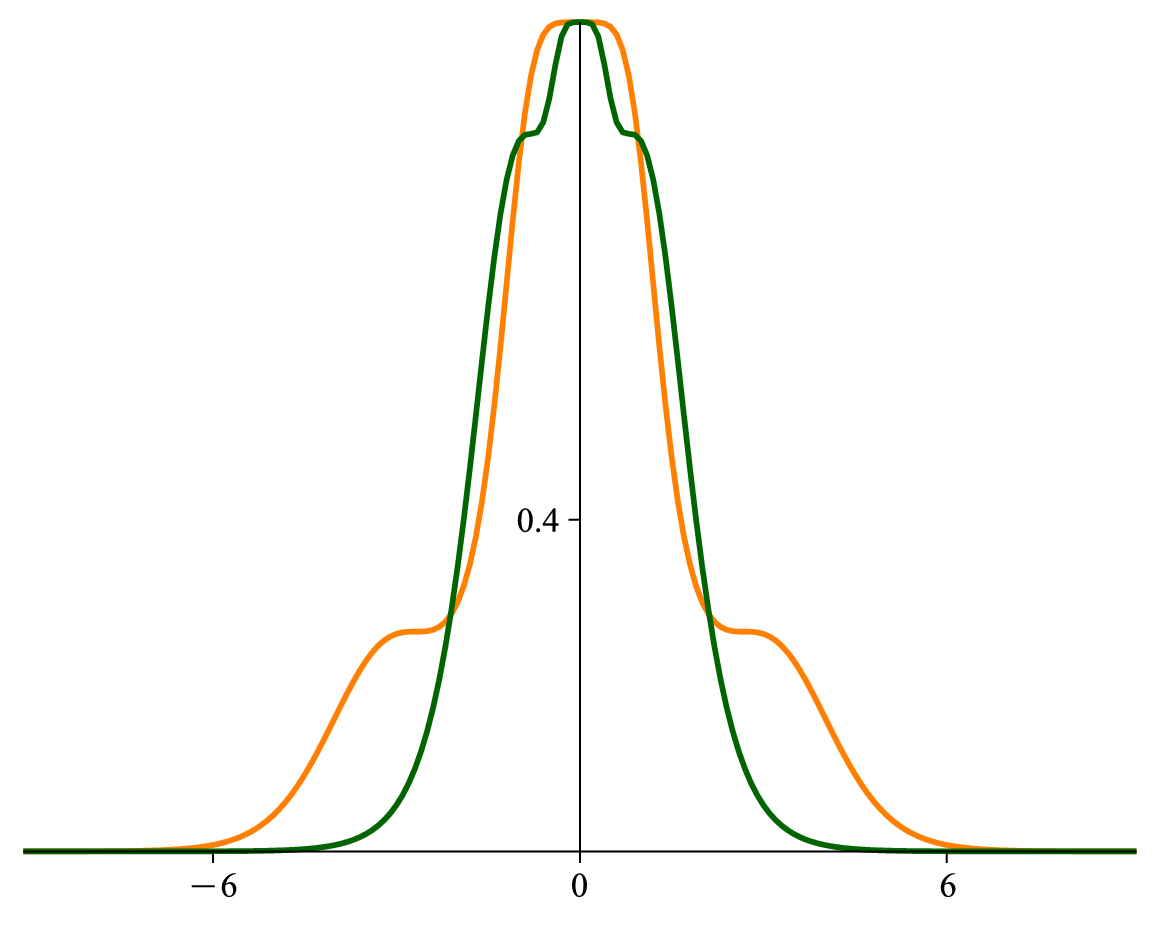}
\includegraphics[width=0.23\textwidth]{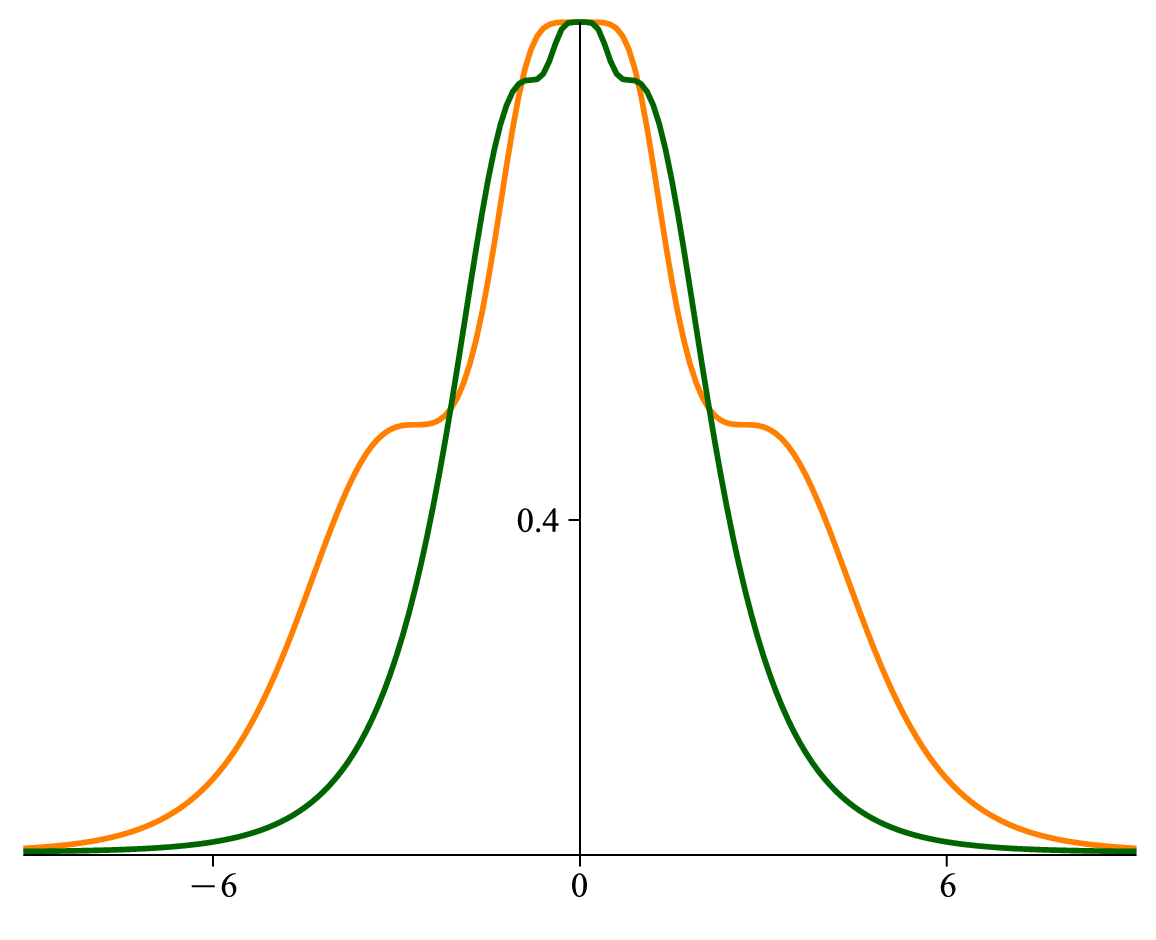}
\includegraphics[width=0.23\textwidth]{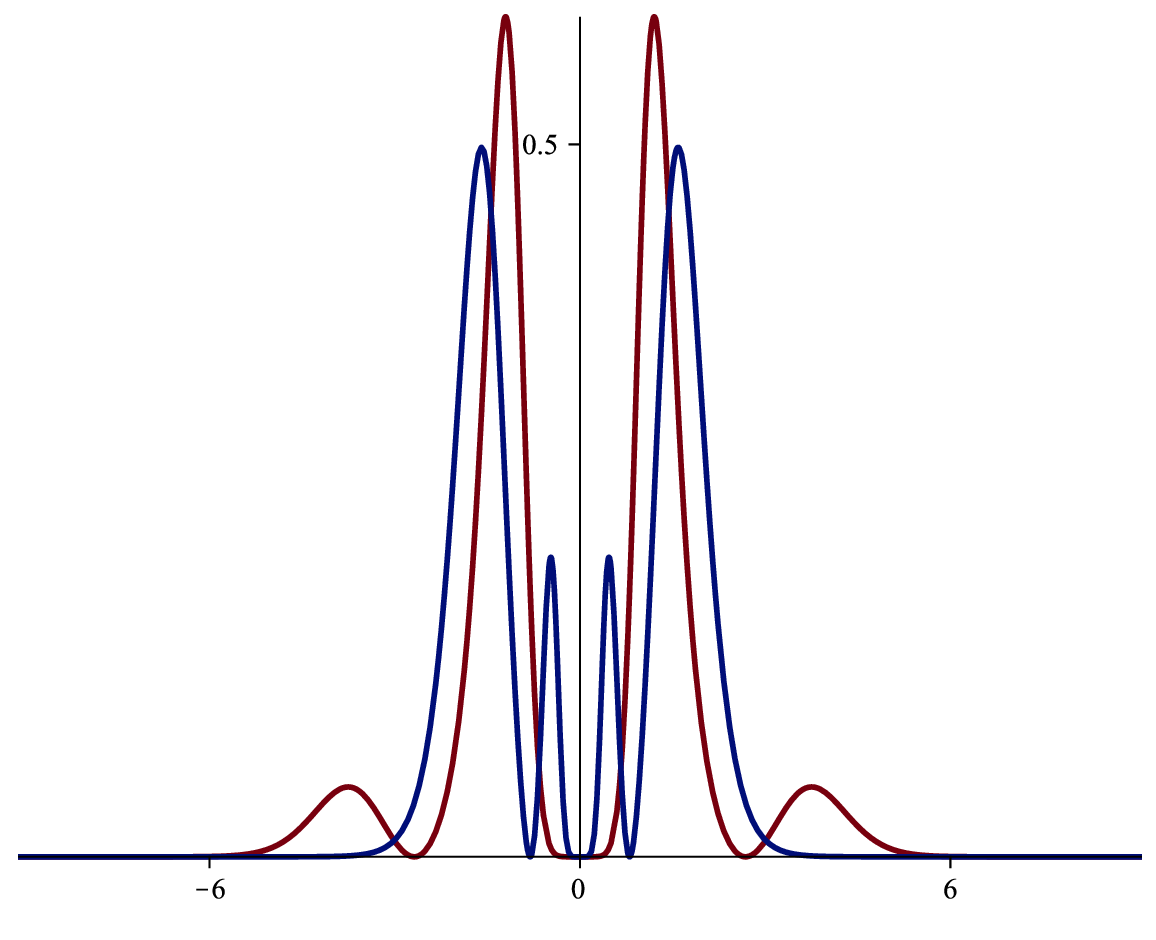}
\includegraphics[width=0.23\textwidth]{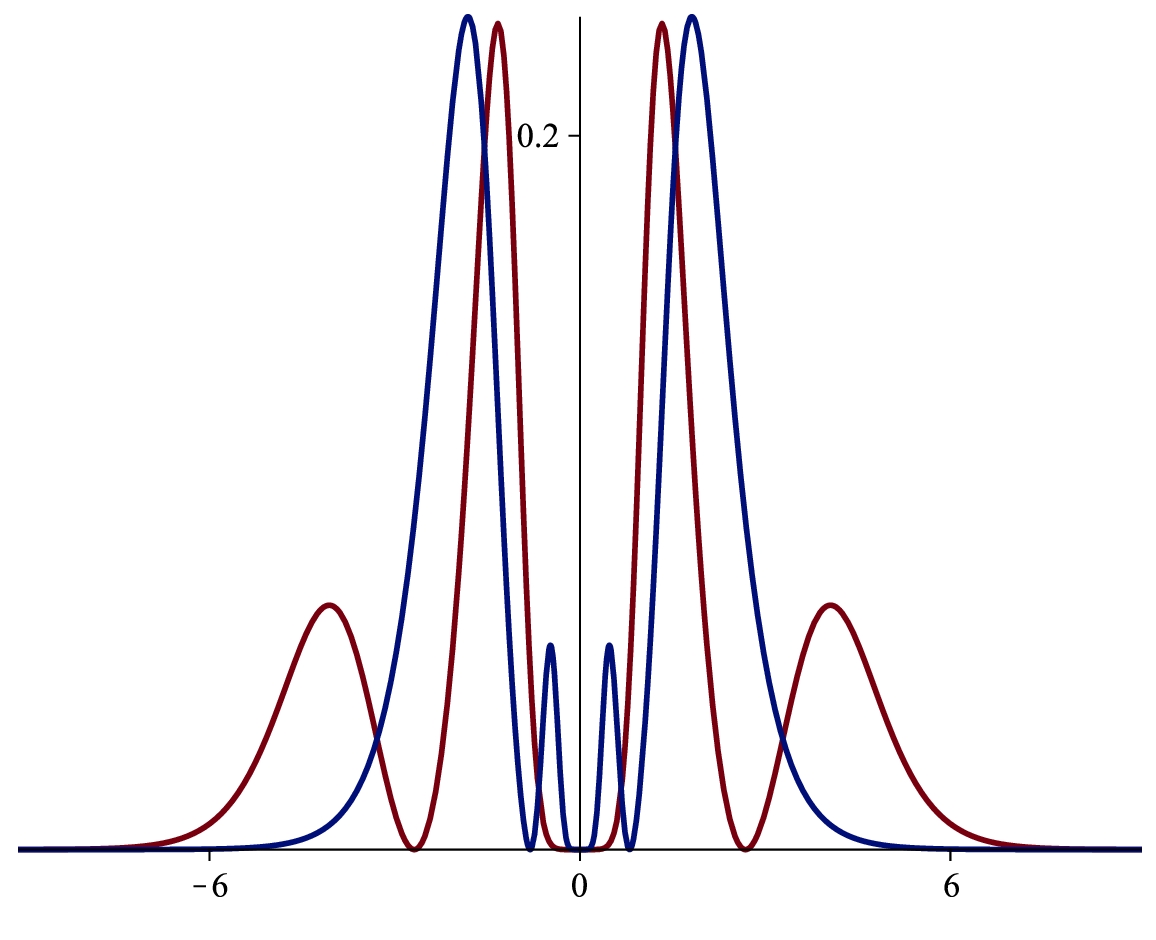}
\caption{Top panel: The lumplike solution $\phi(x)$ in Eq. \eqref{sin}, depicted for $n=1$ (left)  and $2$ (right), for $\alpha=0.3$ (orange) and $1.0$ (green) and for $m=1$. Bottom panel: The energy density $\rho_1(x)$ in Eq. \eqref{ene3}, depicted for $n=1$ (left) and $2$ (right), for $\alpha=0.3$ (red) and $1.0$ (blue), and for $m=1$.}
\label{fig4}
\end{figure}

The energy density $\rho_1$ becomes   
    \begin{eqnarray}
\label{ene3}
        \rho_1(x) = \frac{4}{n^2}\sin^2\left( \left( m+\frac{1}{2}\right) \pi \; \tanh(\alpha x)\right) \nonumber\\
        \times \;{\rm sech}^{4/n}(\eta(x)) \; \tanh^2(\eta(x)).
    \end{eqnarray}
In Fig. \ref{fig4} we depict the solution \eqref{sin} and its corresponding energy density \eqref{ene3} for several values of the parameters involved in the model. The parameter $m$ works in a way similar to the previous case, also controlling the profile of the composed solution, in the form of two-lump, three-lump, etc, as $m$ increases following $m=1,2, \cdots$; one sees that it also works to generate multi-lump configurations.

\begin{figure}
\includegraphics[width=0.23\textwidth]{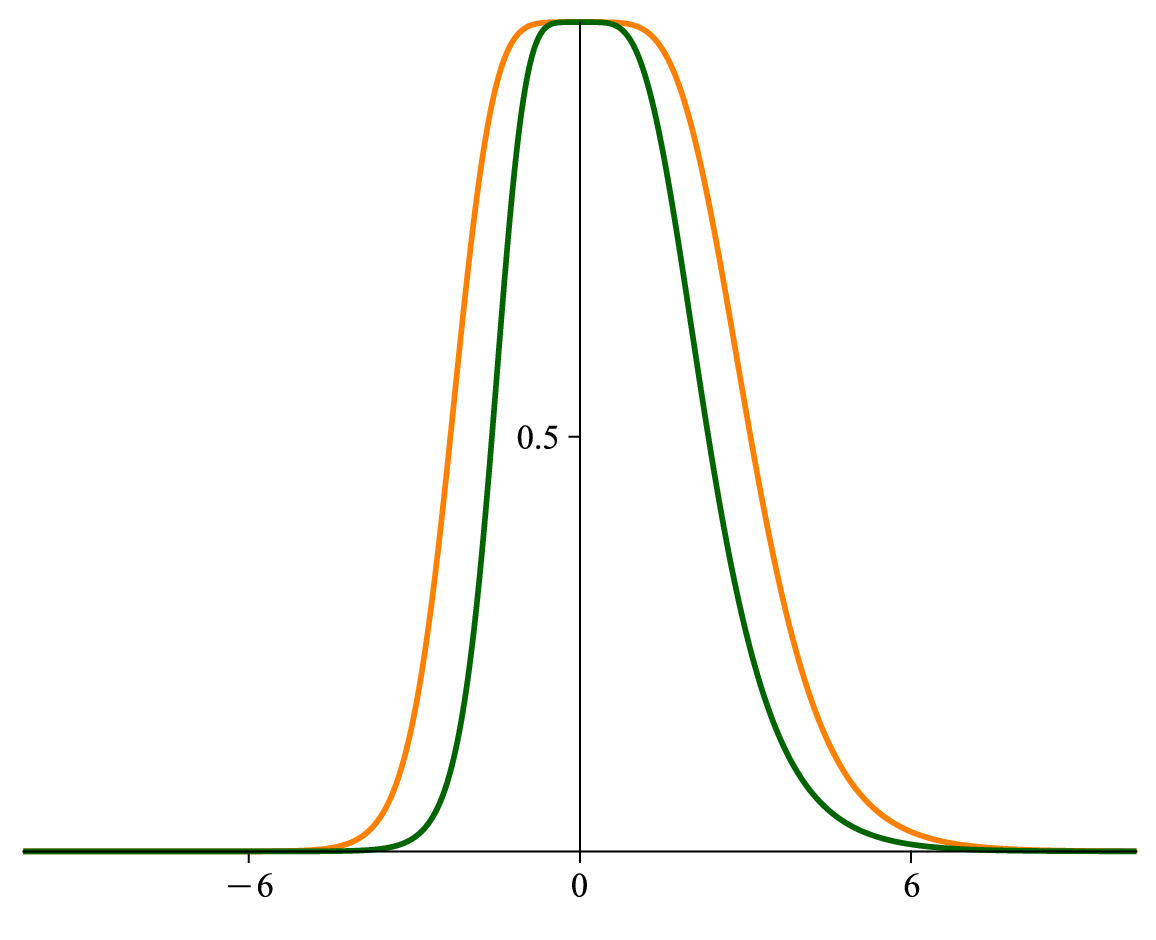}
\includegraphics[width=0.23\textwidth]{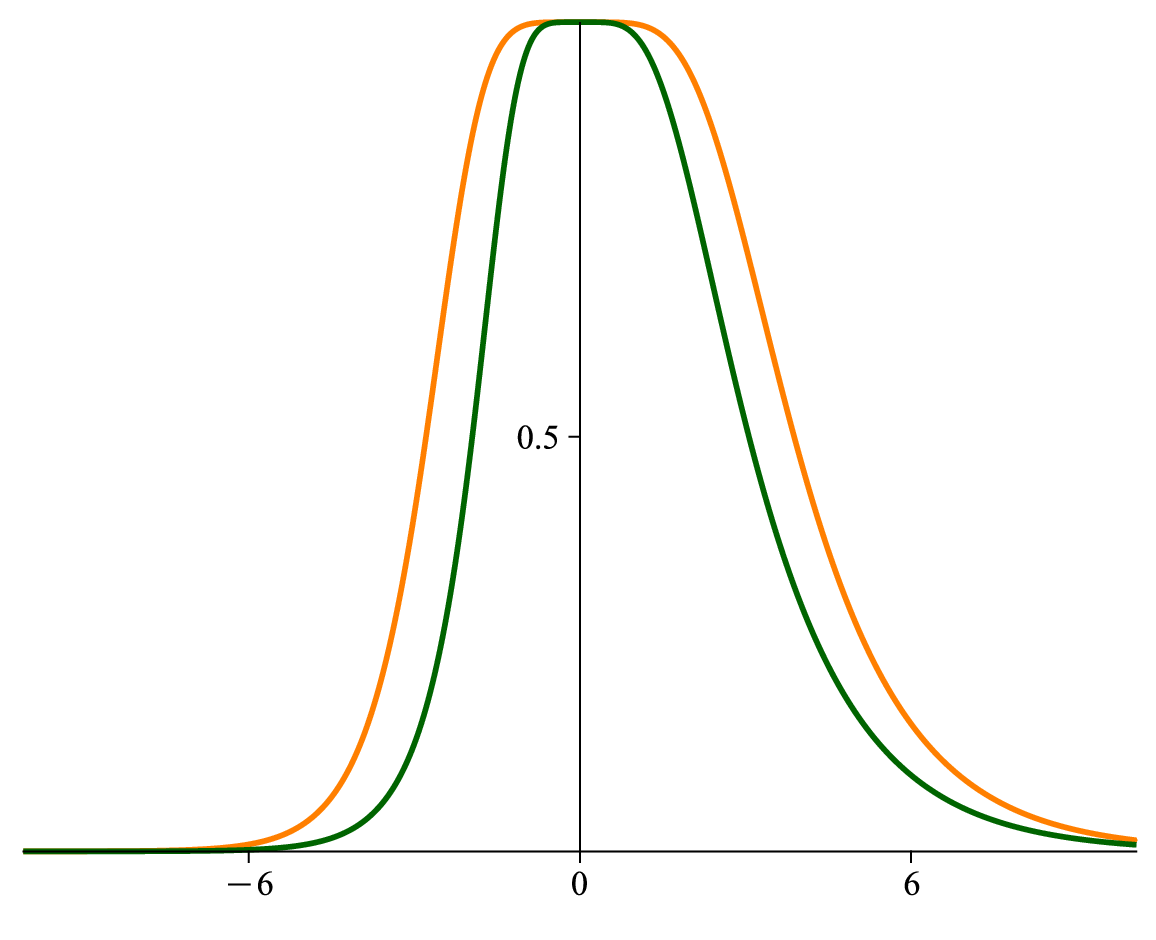}
\includegraphics[width=0.23\textwidth]{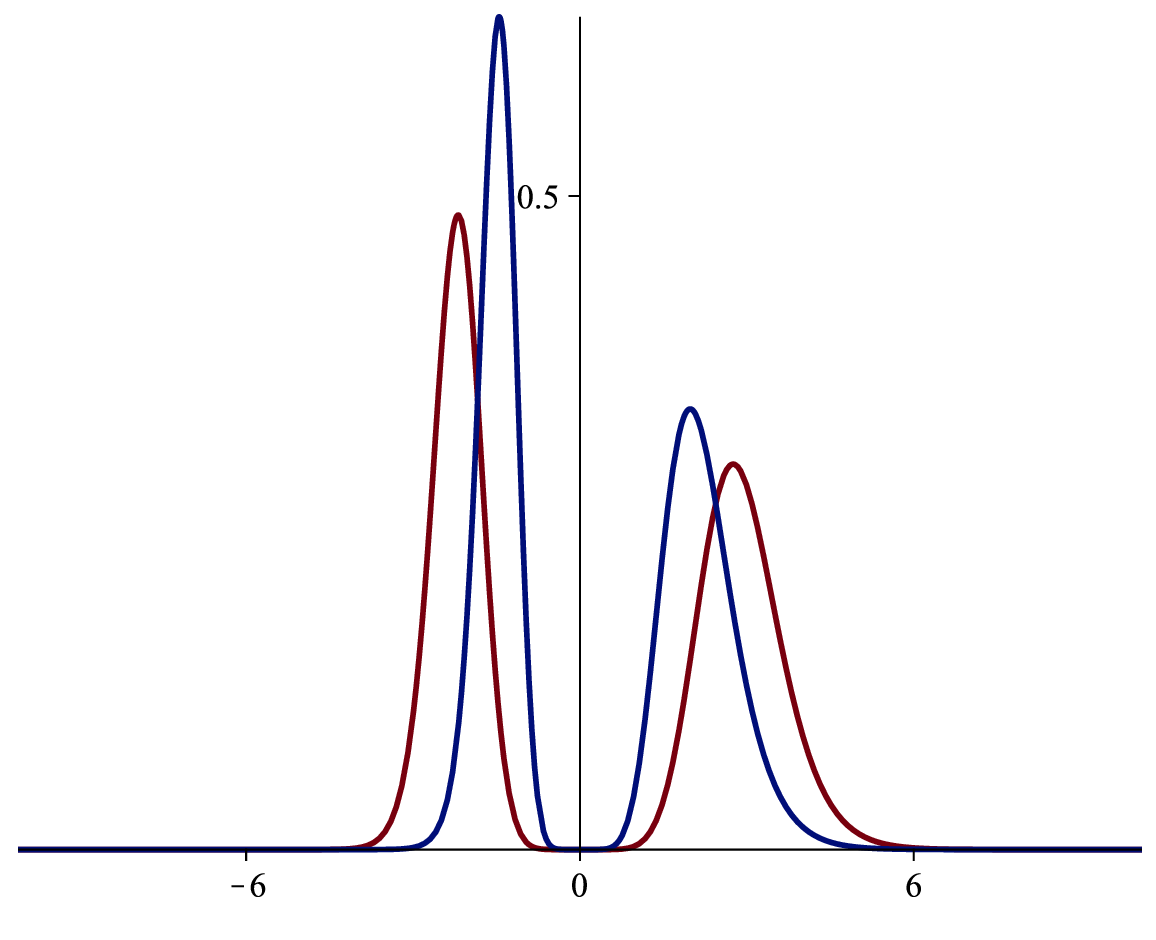}
\includegraphics[width=0.23\textwidth]{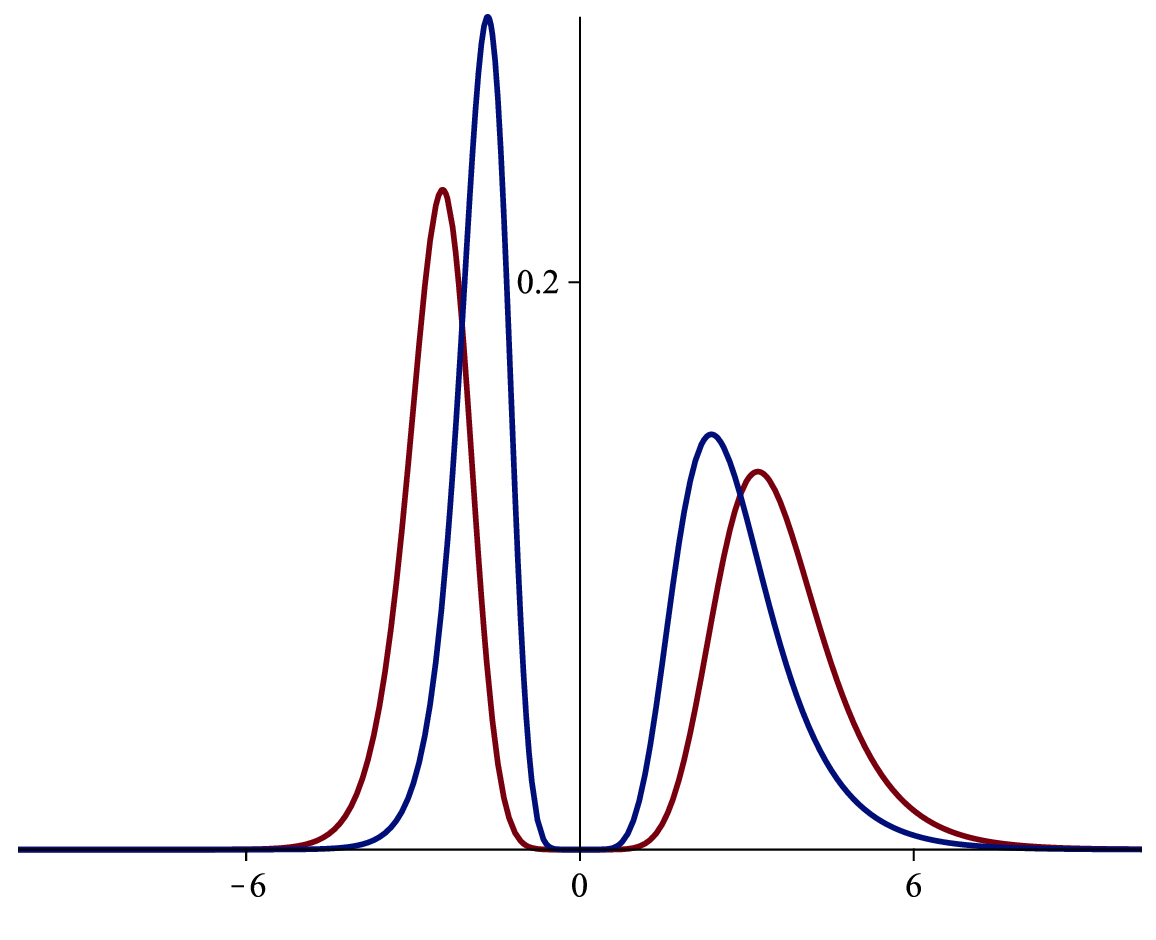}
\caption{Top panel: The lumplike solution $\phi(x)$ in Eq. \eqref{asy}, depicted for $n=1$ (left) and $2$ (right), for $\alpha=0.3$ (orange) and $1.0$ (green) and for $a=0.4$. Bottom panel: The energy density $\rho_1(x)$ in Eq. \eqref{ene4}, depicted for $n=1$ (left) and $2$ (right), for $\alpha=0.3$ (red) and $1.0$ (blue), and for $a=0.4$.}
\label{fig5}
\end{figure}

\begin{figure}[h]
\includegraphics[width=0.23\textwidth]{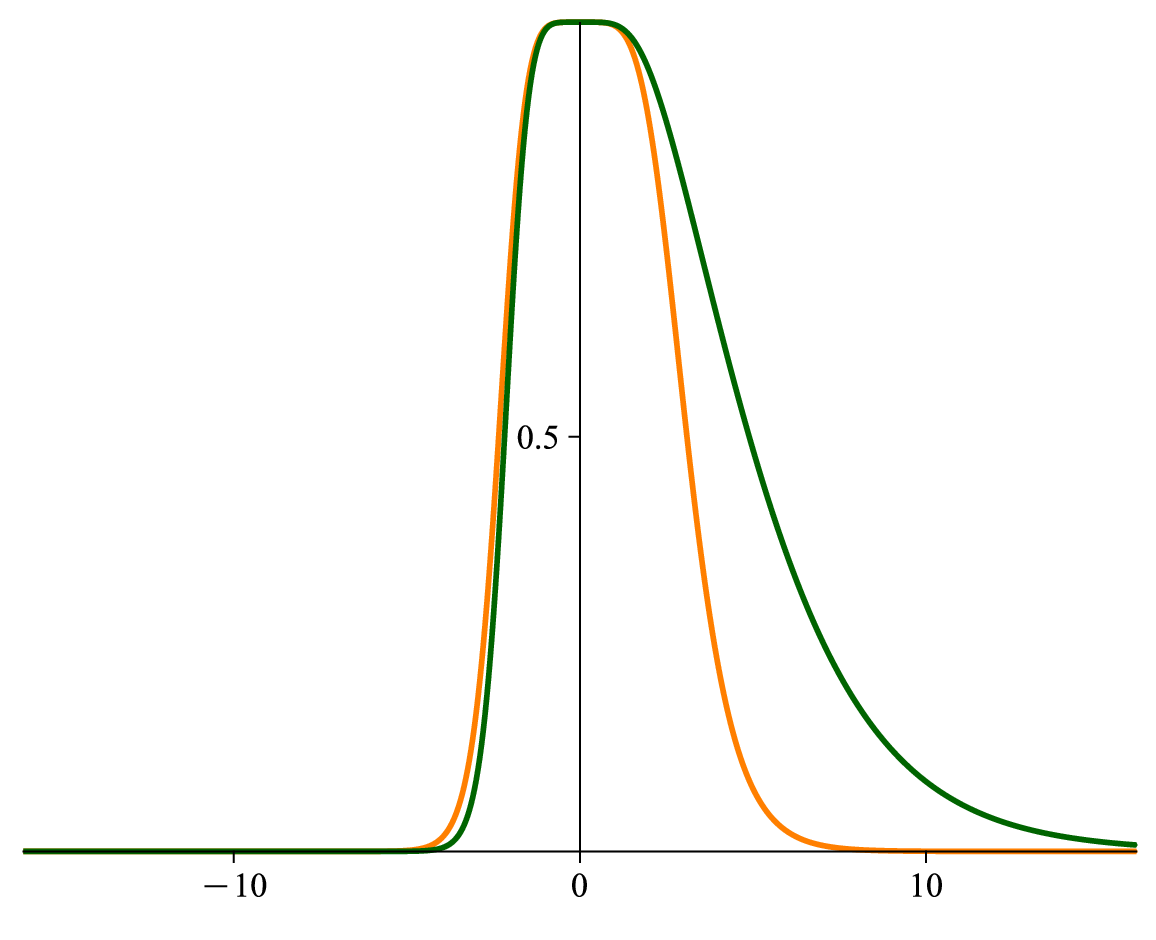}
\includegraphics[width=0.23\textwidth]{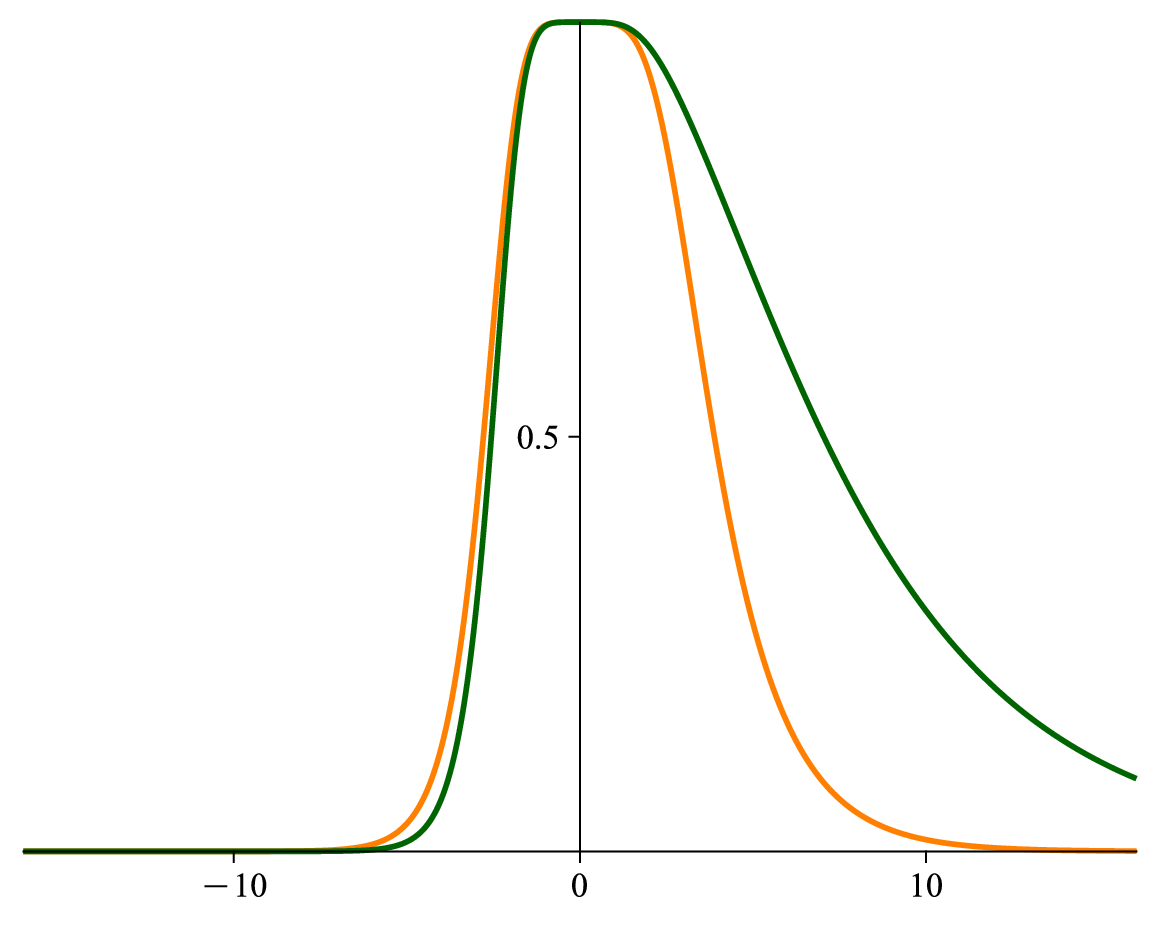}
\includegraphics[width=0.23\textwidth]{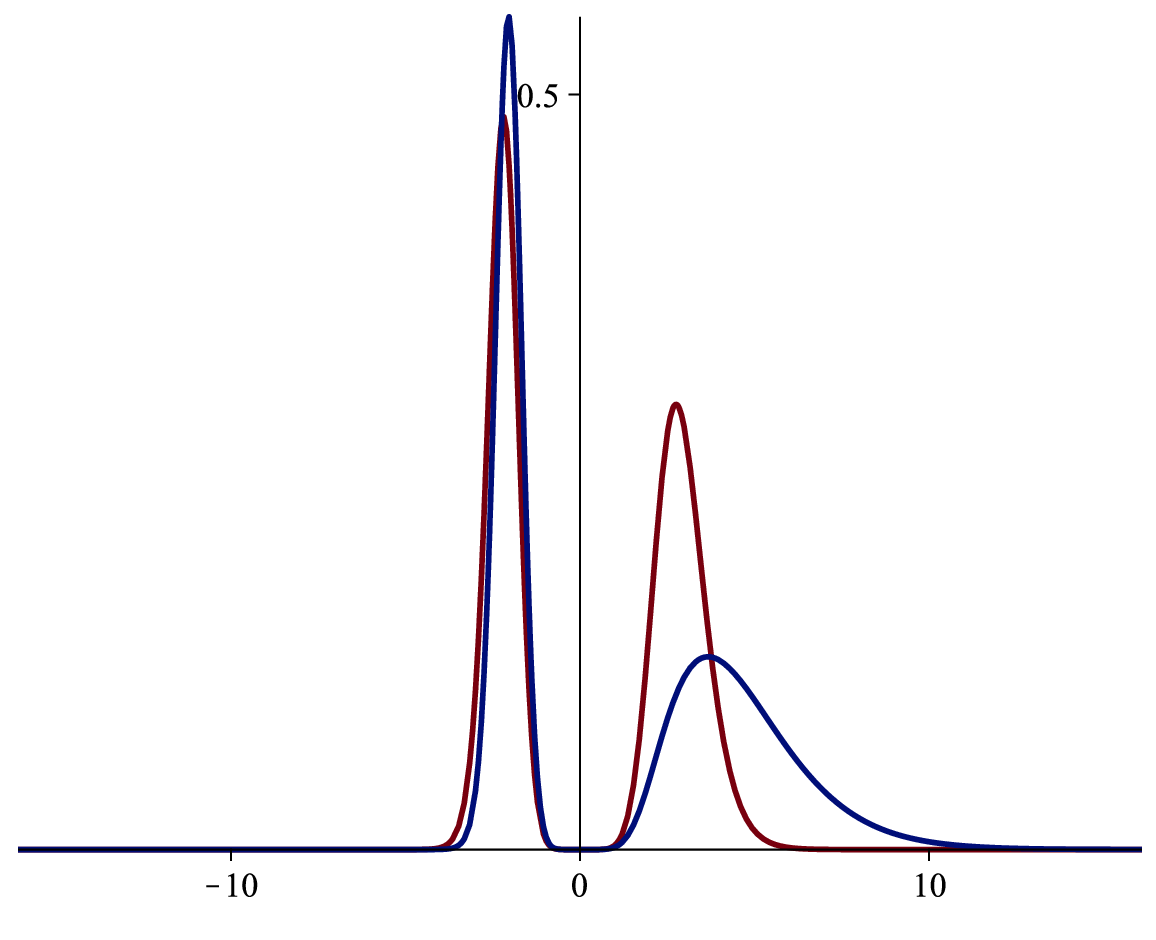}
\includegraphics[width=0.23\textwidth]{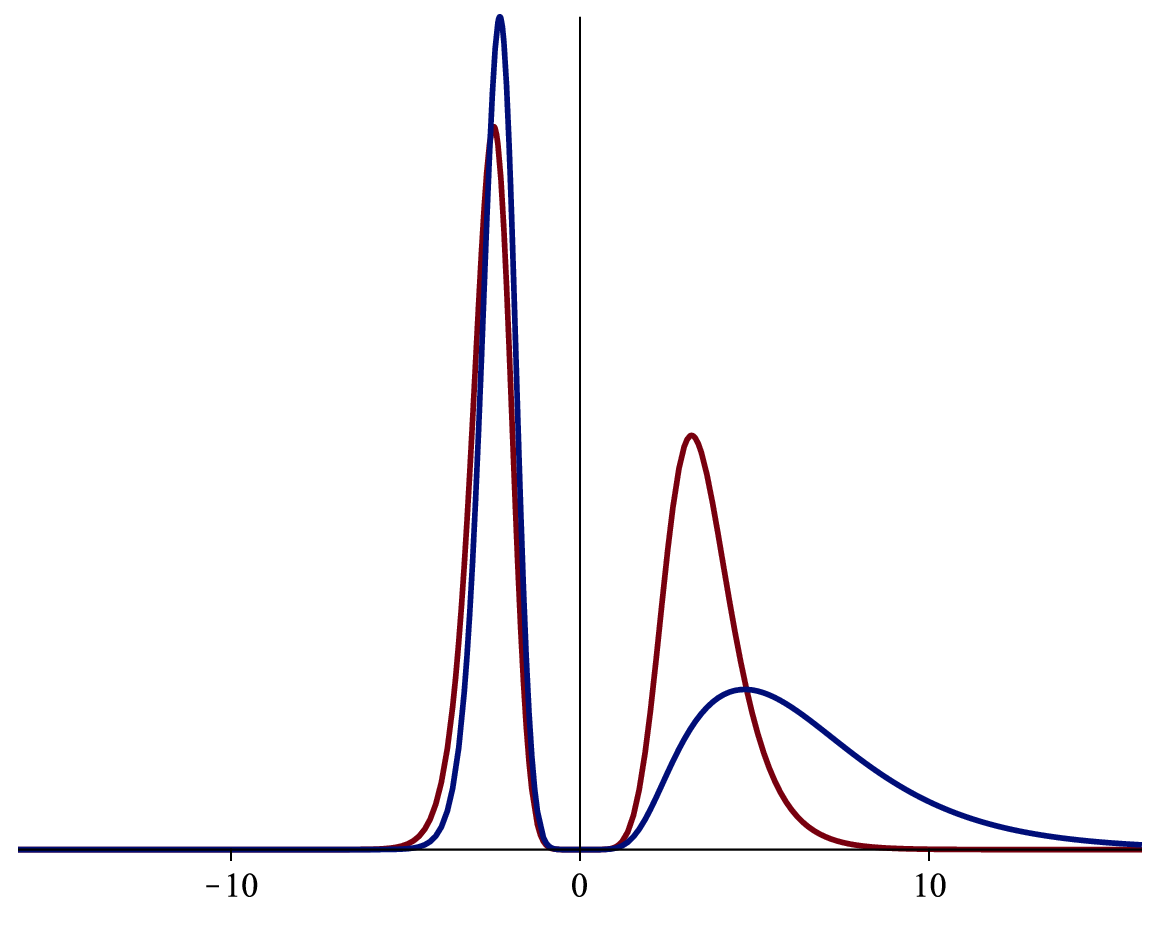}
\caption{Top panel: The lumplike solution $\phi(x)$ in Eq. \eqref{asy}, depicted for $n=1$ (left) and $2$ (right), for $\alpha=0.5$, and for $a=0.4$ (orange) and $0.8$ (green). Bottom panel: The energy density $\rho_1(x)$ in Eq. \eqref{ene4}, depicted for $n=1$ (left) and $2$ (right), for $\alpha=0.5$, and for $a=0.4$ (red) and $0.8$ (blue).}
\label{fig6}
\end{figure}

{\it D. Fourth model. --} We can also consider the function $f(\chi)$ to induce asymmetric behavior in the lumplike solution. To show how this work, let us take $f(\chi)=1/\chi^2(1-a\,\chi)$, with $a$ a real parameter that controls the asymmetry of the solution. Since the configuration $\chi(x)=\tanh(\alpha\,x)$ is inside the interval $[-1,1]$, one has to take $a\in[-1,1]$ to make $f(\chi)$ non negative, as required in the present framework. In this case, the solution is
\be\label{asy}
        \phi(x) = {\rm sech}^{2/n} (\xi (x)),
    \ee
where
    \be \label{eq: coordenada espacial4}
        \xi(x) = x + \frac{a \ln({\rm sech}(\alpha x))}{\alpha} +\frac{\tanh(\alpha x)(a \tanh(\alpha x) - 2)}{2 \alpha},
    \ee 
and the energy density $\rho_1$ is
    \begin{eqnarray} \label{ene4}
        \rho_1(x) &=& \frac{4}{n^2}{\rm sech}^{4/n} (\xi (x))  \,\tanh^2 (\xi (x)) \nonumber\\
        && \times \,\tanh^2 (\alpha x)(1-a \tanh (\alpha x)).
    \end{eqnarray}
The solution and energy density are depicted in Figs. \ref{fig5} and \ref{fig6} for some values of $n$, $\alpha$, and $a$. We noticed that the asymmetry exchanges place for $a\in (0,1]$ and for $a\in[-1,0)$, and in the figures we have used $a=0.4$ and $0.8$ to illustrate how the asymmetry works in this model.

{\bf Conclusions. --} In this work, we followed the procedure described in \cite{G1} to investigate how geometric constrictions described by the presence of the function $f(\chi)$ can be used to induce modifications in a lumplike structure. Since the method works in the presence of first-order differential equations, we also had to follow the work \cite{p2} to implement the present investigation. This was a crucial step, which helped us to deal with first-order differential equations and construct interesting analytical solutions. The procedure was illustrated with several examples, checking all the steps considered to describe the results that appeared in the several figures depicted in the work. The results led us to conclude that the methodology is robust and can be used to explore other models, considering possible applications, in particular, to the study of bright solitons in optical fibers and in Bose-Einstein condensates. The possibility to control the width of the localized configuration, and to generate multi-lump structures 
composed of two or more lumps or asymmetrically distributed around their corresponding centers is certainly of practical use.

There are several other possibilities of continuation of the present study. One of them, in particular, is directly related to the inclusion of electromagnetic and geometric and electromagnetic degrees of freedom, in the so-called Maxwell-scalar and in the Einstein-Maxwell-scalar models, respectively, where interesting solutions may appear in flat  \cite{Her,Baz,Baz2,Mor} and in curved geometries \cite{Herde,Mor2,scal}. Another possibility of current interest is related to applications using the nonlinear Schrödinger equation, following the lines of Refs. \cite{Kono,abc} to describe distinct forms for the resonant bell-shaped bright solitons that appear in such scenarios. Since the methodology described in the present work applies to the case of two scalar fields, for the nonlinear Schrödinger equation we shall naturally need to investigate two coupled equations, but this can be considered following the lines of the two-component Manakov solitons \cite{Man} and also, by \cite{W1} and references therein. The issue here concerns the case of propagation of solitons in nonlinear
coupled waveguides described by coupled nonlinear Schrödinger equations. We think the internal modification of the bright soliton, making it thinner, thicker and also, composed of two or more lumps or asymmetric may be of good use in application of practical use. We are now studying some possibilities, hoping to return to this subject in the near future.

\bigskip

{\bf Acknowledgments:}
This work is partially supported by Conselho Nacional de Desenvolvimento Cient\'\i fico e Tecnol\'ogico (CNPq, Grants 303469/2019-6 (DB), 402830/2023 (DB and RM), 310994/2021-7 (RM)), Coordenação de Aperfeiçoamento de Pessoal de Nível Superior (CAPES, Grant
88887.899549/2023-00 (IB)) and Paraiba State Research Foundation (FAPESQ-PB, Grant 0015/2019).

\end{document}